\documentclass{aa}
\usepackage{graphicx}
\usepackage{txfonts}
\usepackage{color}

\def\eq{\begin{equation}}         
\def\eeq{\end{equation}}      

\begin{document}
\title{Assisted stellar suicide in \object{V617~Sgr}
	\thanks{Based on observations made at Laborat\'orio Nacional de Astrof\'isica/CNPq, Brazil.}
	}

  \author{J. E. Steiner
         \inst{1}
          \and
          A. S. Oliveira \inst{2,3}
	  \and 
	  D. Cieslinski \inst{4}
	  \and
	  T. V. Ricci \inst{1}
          }


   \institute{Instituto de Astronomia, Geof\'{\i}sica e Ci\^encias Atmosf\'ericas, Universidade de S\~ao Paulo, 05508-900,
    S\~ao Paulo, SP, Brasil\\
             \email{steiner@astro.iag.usp.br, tiago@astro.iag.usp.br}
        \and
             SOAR Telescope, Casilla 603, La Serena, Chile\\
	     \email{aoliveira@ctio.noao.edu}
	     \and
	     Laborat\'orio Nacional de Astrof\'{\i}sica / MCT, CP21, Itajub\'a, MG, Brasil
            \and
	    Divis\~ao de Astrof\'{\i}sica, Instituto Nacional de Pesquisas Espaciais, CP 515, S. J. dos Campos, Brasil\\
	    \email{deo@das.inpe.br}
             }

   \date{Received / Accepted}

   \abstract{V617 Sgr is a V Sagittae star -- a group of binaries thought to be the galactic
counterparts of the Compact Binary Supersoft X-ray Sources -- CBSS.}
{To check this hypothesis, we measured the time derivative of its orbital period.}
{Observed timings of eclipse minima spanning over 30,000 orbital cycles are presented.}
{We found that the orbital period evolves quite rapidly: $P/\dot{P} = 1.1\times10^{6}$ years.
This is consistent with the idea that V617 Sgr is a wind driven accretion supersoft source.
  As the binary system evolves with a time-scale of about one million years, which is extremely short
for a low mass evolved binary, it is likely that the system will soon end
either by having its secondary completely evaporated or by the primary
exploding as a supernova of type Ia.}
{}

   \authorrunning{J. E. Steiner et al.}
   \titlerunning{Assisted stellar suicide in \object{V617~Sgr}}
   \maketitle
%

\section{Introduction}

     	Compact Binary Supersoft X-ray Sources (CBSS) are a class of objects that share
in common a set of properties. They are luminous ($\sim$Eddington luminosity)
sources of soft (15-70 eV) X-ray photons and were initially discovered in the
Magellanic Clouds by the Einstein observatory and ROSAT. The CBSS are thought
 to be cataclysmic binaries in which the
secondary is more massive than the primary star. In this situation, when the
secondary fills its Roche lobe a dynamical instability occurs and the mass
transfer takes place on the thermal time-scale, which is about 10 million years
for donor stars of 1--1.5 M$_{\sun}$. This produces accretion rates 100 times larger than
in normal cataclysmic variables and causes hydrostatic nuclear burning on the
surface of the white dwarf (see Kahabka \& van den Heuvel \cite{kah} for a
review). 

Only two CBSS (\object{MR Vel} and \object{QR And}) are found in the Galaxy, where one should find about a thousand.
This is presumably due to the absorption of their soft X-ray emission by the interstellar gas 
in the Galactic plane.
V Sagittae stars (Steiner \& Diaz \cite{stei98}) were proposed
as a new class of binaries that display properties quite similar to those of
CBSS, but are not detected as supersoft sources. They may be the galactic
counterpart of the CBSS. The soft photons are either absorbed by the stellar
wind or by the interstellar medium (or both). In case this hypothesis is
correct, these two classes should share a number of properties in common. For
example, the time variation of the orbital period should be high and similar in
the two situations  - and this could be a critical test for the hypothesis of the
CBSS -- V~Sge connection.

What do we expect in terms of the orbital period time derivative? In the scenario of dynamical instability,
we expect that the orbital period decreases with time. There is only one such
object for which the period derivative has been measured: \object{V~Sge}. Its period, in
fact, decreases with a time-scale of 5 million years (Patterson et al. \cite{patter}). However, this scenario
only predicts the existence of orbital periods longer than 6 hours
(Deutschmann \cite{deutsch}; King et al. \cite{king2}). For periods smaller than this limit, the
mass transfer is too small for nuclear burning to occur.
This limitation on the orbital period imposes a problem to the interpretation of the short orbital period 
systems among CBSS (\object{SMC~13} and \object{RX J0537.7-7034}) and among V~Sge stars (\object{V617~Sgr}), which
have orbital periods shorter than 5 hours.

It has been proposed that in such objects the mass transfer occurs because of
radiation induced wind from the secondary (van Teeseling \& King \cite{tees}). If a low
mass secondary star ($M_{2}\leq0.6 M_{\sun}$) looses mass adiabatically, that is, on a
time-scale short compared to the thermal time-scale, it expands as 

\eq
\frac{\dot{R_{2}}}{R_{2}} = \zeta \frac{\dot{M_{2}}}{M_{2}}
\eeq

\noindent where $\zeta=-1/3$ is the effective mass-radius index of the secondary star.  In this
case, the expansion forces mass transfer through the internal Lagrangean point
at a high rate, producing a CBSS. In this situation, the orbital period
increases with time (van Teeseling \& King \cite{tees}). 

We have, thus, two distinct scenarios for CBSS as well as for V~Sge stars:
systems with initial orbital periods longer than 6 hours evolve with negative
derivative while systems with shorter periods should evolve with positive
derivatives, both with high absolute values. For wind driven supersoft
binaries, the orbital period increases with time and may become longer than 6
hours. The total possible orbital period range is 2--30 hours (van Teeseling 
\& King \cite{tees}). So, for a given object, how to decide to which of the two above
paradigms the object belongs? In principle one could determine the mass ratio 
or the secondary's spectral type. 
But no secondary star has been detected so far in any of the CBSS
or V~Sge stars. Emission lines are strongly contaminated by the wind and are not
reliable dynamical indicators.

For eclipsing systems there is an alternative way of finding
whether a given star belongs to one class or to the other: by measuring its orbital period time
derivative. A critical test could be provided by V617~Sgr. A positive period
derivative with time-scale of about a million years is predicted. This should
be measurable, given the time base of our eclipse timings. A similar study was
proposed for \object{T~Pyx}, a recurrent nova with an orbital period of 1.8 hour
(Knigge et al. \cite{knigge}) .

V617~Sgr (Steiner et al. \cite{stei99}; Cieslinski et al. \cite{cies}) was identified as a V~Sge
star (Steiner \& Diaz \cite{stei98}) with an orbital period of 4.98 hours. It presents
a light curve with two maxima and two minima (see Fig. 3 in Steiner et al.
\cite{stei99}) of unequal depths. The system presents high and low photometric states,
like V~Sge itself and most CBSS. Timings of the main (deepest) minima provide
an opportunity to measure the orbital period with accuracy as well as its time
derivative.

\section{Observations and data analysis}

V617~Sgr was initially classified as a Wolf-Rayet star and received the
designation of WR 109 (van der Hucht et al. \cite{huch1}). It was also classified as an
irregular variable of type I in the General Catalogue of Variable Stars
(Kholopov et al. \cite{kholo}) and, for that reason, included in a program to search
for new close binary systems (Steiner et al. \cite{stei88}) when its orbital period was
discovered. Since then, timings for the light curve minima of V617~Sgr were
measured with distinct telescopes, photometers and CCDs at the Observat\'orio
Pico dos Dias -- LNA/MCT -- in southeast Brazil, 
spanning an interval of more than 32,000 orbital cycles.

We obtained new observations of V617~Sgr with the Zeiss 60 cm telescope of OPD
in 2003, 2004 and 2005. We employed a thin, back-illuminated EEV CCD 002-06 chip and
a Wright Instruments thermoelectrically cooled camera. The images were obtained
through the Johnson V filter and were corrected for bias and flat-field, using
the standard IRAF~\footnote
{IRAF is distributed by the National Optical Astronomy Observatories,
which are operated by the Association of Universities for Research in Astronomy, Inc., under cooperative agreement
with the National Science Foundation.} routines. The differential aperture photometry was
performed with the aid of the DAOPHOT II package. In Table~\ref{timings} we list the
timings of minimum light of V617~Sgr. This list includes the photometric minima
already presented in Steiner et al. (\cite{stei99}) along with 17 new timings determined
in this work. 

\begin{figure}
      \resizebox{\hsize}{!}{\includegraphics{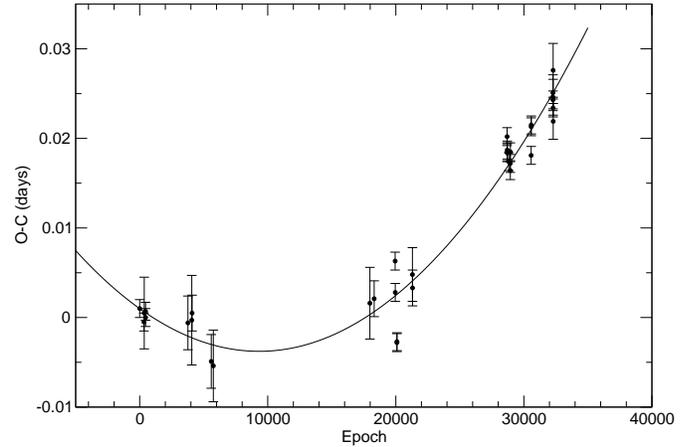}}
     \caption{$O-C$ diagram of the timings of minima observed from 1987 to 2005, relative to the constant period
     of 0.2071667 days from the ephemeris given by Steiner et al. (\cite{stei99}). The fitted parabola indicates 
     a period that increases with  $\dot{P} = 1.1\times10^{-10}$ day/cycle.}
     \label{oc}
   \end{figure}

An $O-C$ diagram constructed for all timings relative to the linear ephemeris
given by Steiner et al. (\cite{stei99}) is shown in Fig.~\ref{oc}. A quadratic function fits
the data and yields the ephemeris

\begin{eqnarray}
\mathrm{Minimum~light}  = \mathrm{HJD}~~2,446,878.7730(\pm7) 
\nonumber\\
 +0.20716568(\pm4) \times E + 5.5(\pm1)\times10^{-11} \times E^2
\end{eqnarray}

This implies $\dot{P} = 1.1\times10^{-10}$ day/cycle, so the observed time-scale of period
change is $1.1\times10^{6}$ year. 
 A first point to notice is that the time derivative is positive, as expected in a
system that follows the paradigm of a wind driven accretion supersoft X-ray
source. Second, this time-scale is much shorter than the one of V~Sge, which
has a negative $\dot{P}$ and has a time-scale of $5\times10^6$ year (Patterson et al. \cite{patter}), almost 5 times
longer than that of V617~Sgr.

\begin{table*}[!h]
\caption{Times of minima.}
\label{timings}
\begin{flushleft}
\begin{tabular}{ll}
\hline
\hline\noalign{\smallskip\smallskip}
            \noalign{\smallskip}
 Obs. minimum   &E\\
 (HJD)   & (cycle)\\
\hline
\noalign{\smallskip}
2446878.773(1)	& 0 \\
2446947.758(1)	& 333 \\
2446952.731(4)	& 357 \\
2446973.655(1)	& 458 \\
2446974.483(1)	& 462 \\
2447658.754(3)	& 3765 \\
2447721.733(5)	& 4069 \\
2447725.670(2)	& 4088 \\
2448036.829(3)	& 5590 \\
2448069.768(4)	& 5749 \\
2450246.685(6)	& 16257\\
2450602.595(4)	& 17975\\
2450671.582(2)	& 18308\\
2451011.754(1)	& 19950\\
2451013.615(1)	& 19959\\
2451040.541(1)	& 20089\\
2451041.577(1)	& 20094\\
2451290.806(3)	& 21297\\
2451292.669(2)	& 21306\\
2452822.612(1)	& 28691\\
2452823.646(1)	& 28696\\
2452824.682(1)	& 28701\\
2452825.718(1)	& 28706\\
2452849.748(1)	& 28822\\
2452873.572(1)	& 28937\\
2452874.607(1)	& 28942\\
2452875.645(1)	& 28947\\
2453211.669(1)	& 30569\\
2453212.708(1)	& 30574\\
2453213.744(1)	& 30579\\
2453564.688(2)	& 32273\\
2453565.723(1)	& 32278\\
2453566.552(2)	& 32282\\
2453566.758(1)	& 32283\\
2453567.798(3)	& 32288\\
2453568.621(2)	& 32292\\
\noalign{\smallskip}
\hline
\end{tabular}
\end{flushleft}
\smallskip\noindent
\end{table*}

\section{Discussion and conclusions} 

An important system parameter is the mass of the white dwarf. This is usually
derived from dynamical measurements. In the case of this system, however, the
observed strong wind complicates this kind of determination (Cieslinski et al. \cite{cies}).
Steiner et al. (\cite{stei99}) considered that the mass of the white dwarf must be
quite low. This, however, was based on the erroneous assumption that the system
is a 
CBSS with a secondary more massive than the primary and that accretion is driven by the thermal expansion of the donor star.  

One alternative way of estimating this mass is by measuring the time-scale of
decline from outburst maxima (Southwell et al. \cite{south}). V617~Sgr displays such
outbursts as most of other well observed V~Sge and CBSS systems do. One such
event can be seen in Fig. 4 of Steiner et al. (\cite{stei99}). Eclipse cycle 4069 was
observed when the system was at high state. Four days later cycle 4088 was
observed at about one magnitude fainter. The decay timescale is, therefore, 4
days or shorter.

From the outburst decline we derive a high mass for the white dwarf: $M_1 = 1.2
M_{\sun}$ or higher. If one assumes that the secondary is in the main sequence (this
is not obvious, as it is not necessarily in equilibrium), $M_2 = 0.48 M_{\sun}$ and the 
mass ratio is $q= 0.40$.

In the wind driven supersoft X-ray binary scenario, the orbital period time
derivative is given by the formula (Knigge et al. \cite{knigge})

\eq
\frac{P}{\dot{P}} = \frac{2M_2}{(3\zeta -1)(1+g)\dot{M}_{w2}}
\eeq

\noindent where 

\eq
\dot{M}_{acc}= -g \dot{M}_{w2}
\simeq 1.2 \times 10^{-6} g^{2} \phi^{2} \eta_a \eta_s \left(\frac{q^{5/2}}{1+q}\right)^{2/3} m_1~M_{\sun} yr^{-1}
\eeq

\noindent and 

\eq
g=\frac{(6\beta_2 + 2q)-(5+3\zeta)(1+q)}{(1+q)(5+3\zeta-6q)}
\eeq

here $q=M_2/M_1$ is the mass ratio, $\dot{M}_{w2}$ is the wind mass loss rate 
from the irradiated secondary star, $\dot{M}_{acc}$ is the accretion rate, $m_1=M_{1}/M_{\sun}$,
$\eta_a$ measures the luminosity per gram of matter accreted relative to the value
for nuclear shell burning,
$\eta_s \sim 1$ for CBSS (the efficiency of the primary's spectrum in producing ionizing
photons and driving a wind),
$\beta_2$ is the specific angular momentum loss of the secondary star,
and $\phi$ is an efficiency factor parameterizing the fraction of the companion's
face which is irradiated. 

In the present case the period derivative and the mass ratio are self-consistent if $\beta_2 = 1.2$. That
is the ratio of the  specific momentum of the wind relative to that of the
secondary star. The observed period variation suggests that V617~Sgr is,
indeed, a wind-driven supersoft X-ray source. This adds more evidence to the
hypothesis that V~Sge stars are the galactic counterparts of the CBSS objects. 

This binary system evolves with a time-scale of about one million years,
which is extremely short for this kind of low-mass evolved binary. As the white dwarf is quite 
massive and if it accretes half of the mass of the
secondary star it may soon reach the Chandrasekhar limit and, eventually, explode as a
Supernovae type Ia. Such binary systems may
occur in old populations as demanded by SN Ia statistics, contrary to thermal
time-scale mass transfer systems that require relatively young populations.
The other possibility is that the secondary will evaporate completely.
In this
situation one would expect the orbital periods to increase up to 30 hr (van Teeseling
\& King \cite{tees}).

V617~Sgr is a system that has left completely the standard CV evolutionary
track and will probably destroy itself. This is a channel that may remove this
kind of system from the general CV population.

V617~Sgr is a member of a group of 3 known stars that are wind driven
supersoft x-ray binaries; the other members of the group are SMC~13 and T~Pyx 
(van Teeseling \& King \cite{tees}; Knigge et al. \cite{knigge}).
It is the first system in which the predicted evolutionary trend has been
clearly observed, thanks to the existence of a relatively deep eclipse in the
orbital light curve.

\begin{acknowledgements}
We acknowledge Barbara Kato for help in observations and data reduction.
A. S. Oliveira and T. V. Ricci thank FAPESP for financial support.
\end{acknowledgements}


\end{document}